\documentclass[3p,times,procedia]{elsarticle}
\usepackage{nupha_ecrc}
\usepackage{lineno}
\usepackage{graphicx}
\usepackage[loose]{units}
\usepackage{upgreek}
\usepackage{color}
\usepackage[usenames,dvipsnames]{xcolor}
\usepackage{amsmath,amssymb,amsbsy}
			
\newcommand{\pT}{p_{\mbox{\tiny T}}}
\newcommand{\sNN}{$\sqrt{s_{\mbox{\tiny NN}}}$ }
\newcommand{\s}{$\sqrt{s}$ }

\newcommand{\Pb}{{\mbox{Pb--Pb}} }

\newcommand{\pp}{pp }

\volume{00}

\firstpage{1}

\journalname{Nuclear Physics A}

\runauth{}

\jid{nupha}

\jnltitlelogo{Nuclear Physics A}

\usepackage{amssymb}

\usepackage[figuresright]{rotating}

\begin{document}

\begin{frontmatter}



\dochead{XXVIIIth International Conference on Ultrarelativistic Nucleus-Nucleus Collisions\\ (Quark Matter 2019)}

\title{Light neutral meson production in heavy ion collisions with ALICE in the era of precision physics at the LHC}


\author{Mike Sas (mike.sas@cern.ch), for the ALICE Collaboration}

\address{University of Utrecht \& Nikhef, Netherlands}

\begin{abstract}
The production of light neutral mesons in AA collisions probes the physics of the Quark-Gluon Plasma (QGP), which is formed in heavy-ion collisions at the LHC. More specifically, the centrality dependent neutral meson spectra in AA collisions compared to its spectra in minimum-bias pp collisions,  scaled with the number of hard collisions, provides information on the energy loss of partons traversing the QGP. The measurement allows to test with high precision the predictions of theoretical model calculations. In addition, the decay of the $\pi^{0}$ and $\eta$ mesons are the dominant backgrounds for all direct photon measurements. Therefore, pushing the limits of the precision of neutral meson production is key to learning about the temperature and space-time evolution of the QGP.

In the ALICE experiment neutral mesons can be detected via their decay into two photons. The latter can be reconstructed using the two calorimeters EMCal and PHOS or via conversions in the detector material. The excellent momentum resolution of the conversion photons down to very low $\pT$ and the high reconstruction efficiency and triggering capability of calorimeters at high $\pT$, allow us to measure the $\pT$ dependent invariant yield of light neutral mesons over a wide kinematic range.

Combining state-of-the-art reconstruction techniques with the high statistics delivered by the LHC in Run 2 gives us the opportunity to enhance the precision of our measurements. In these proceedings, new ALICE run 2 preliminary results for neutral meson production in \pp and \Pb collisions at LHC energies are presented.

\end{abstract}

\begin{keyword}
ALICE \sep neutral mesons \sep neutral pion \sep eta meson \sep nuclear modification

\end{keyword}

\end{frontmatter}


\section{Introduction}

In the field of heavy-ion physics we are faced with fundamental questions: What are the different particle production mechanisms across different system sizes? Can we find the onset of the QGP in heavy-ion collisions? Is there a QGP droplet formed in small collision systems~\cite{Wilke:2018}?
In proton--proton collisions the particle production mechanism at high $\pT$ $(\gtrsim 6$ GeV/$c)$ is expected to be dominated by the fragmentation of high momentum partons in jet-like structures. In collisions of heavy nuclei such as \Pb, the production of particles is expected to be dominated by the hadronisation of the QGP for low $\pT$ $(\lesssim 6$ GeV/$c)$, while modification of the hadron production at higher $\pT$ will also be influenced by parton-QGP interactions. Studying the particle production mechanisms is thus key to understand the physics governing both small and large systems.

Identified hadron spectra are a good probe to study both the production mechanisms in pp collisions~\cite{Sassot:2010}, as well as the parton energy loss in high-energy heavy-ion collisions. Among these identified hadrons are the neutral pion ($\pi^{0}$) and $\eta$ meson, which are abundant and have large branching ratios into two photons, making them suitable probes to study details of particle production. In addition, measuring neutral mesons grants the possibility of extracting the direct photon signal that is seen as an excess yield above the photons from hadronic decays, and is probing e.g. the temperature of the QGP~\cite{Shen:2013vja}.

We present the invariant yield of the $\pi^{0}$ and $\eta$ mesons in pp at \s$ = 5$ TeV and \Pb collisions at\\ \sNN$ = 5.02$ TeV, as measured with the ALICE detector.

\section{Method}
\subsection{Photon reconstruction}
The photons are reconstructed with the photon conversion method (PCM) and the calorimeters PHOS and EMCal.
With PCM, the photons that convert in the detector material to an electron-positron pair are reconstructed using the ITS and TPC detectors. With a conversion probability of 8\%, this method is limited in statistical precision but it profits greatly from the high momentum resolution of the ALICE central barrel.\\
The calorimeters are situated outside the inner detectors and are able to measure the photons by absorbing their full energy in their calorimeter towers.
The PHOS calorimeter consists of lead tungstate crystals with a cell size of 2.2 cm $\times$ 2.2 cm installed at a radius of 4.6 m from the collision region.
The EMCal is a Pb-scintillator sampling calorimeter with a cell size of 6 cm $\times$ 6 cm installed at a radius of 4.28 m from the collision region. The PHOS has a better energy resolution but the EMCal has a larger acceptance. Both calorimeters have trigger capabilities to select events with high energetic clusters.

\subsection{Meson reconstruction}
The neutral mesons are reconstructed as follows.
First, the photons are reconstructed and the invariant mass of every photon pair is calculated. Second, the meson raw yield is obtained by integrating the invariant mass distributions around its corresponding mass, after subtracting the combinatorial and remaining background. Third, the raw yield is corrected for efficiency, acceptance, and feed-down from secondaries. Analyses were performed for various combinations of photons (PCM-PCM, PCM-PHOS, etc.) and then combined to a single result. Combining these different photon reconstruction techniques allows us to reduce the statistical and systematic uncertainties of the neutral meson spectra.

\begin{figure}
	\centering
	\includegraphics[width=0.49\textwidth]{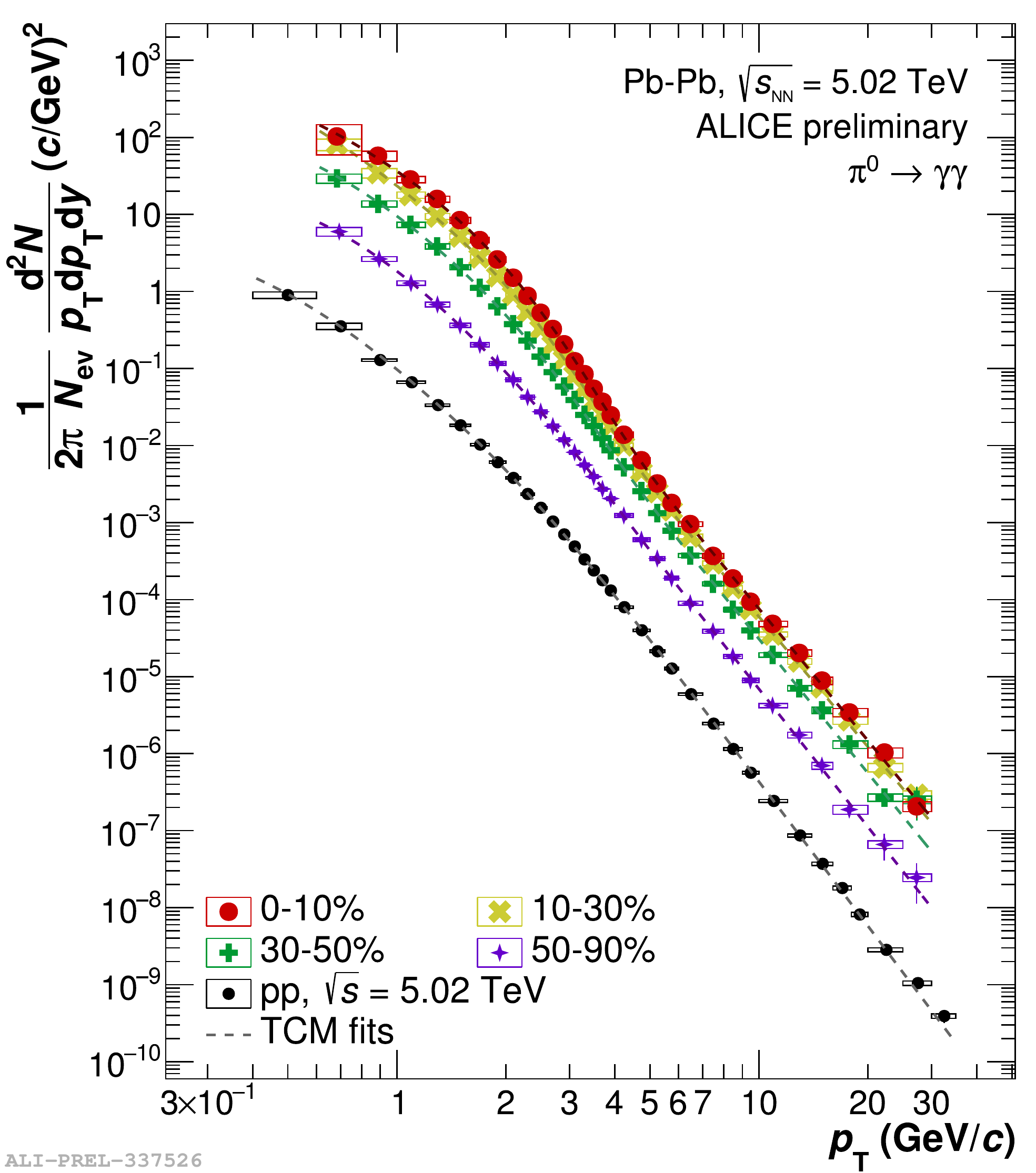}
	\includegraphics[width=0.49\textwidth]{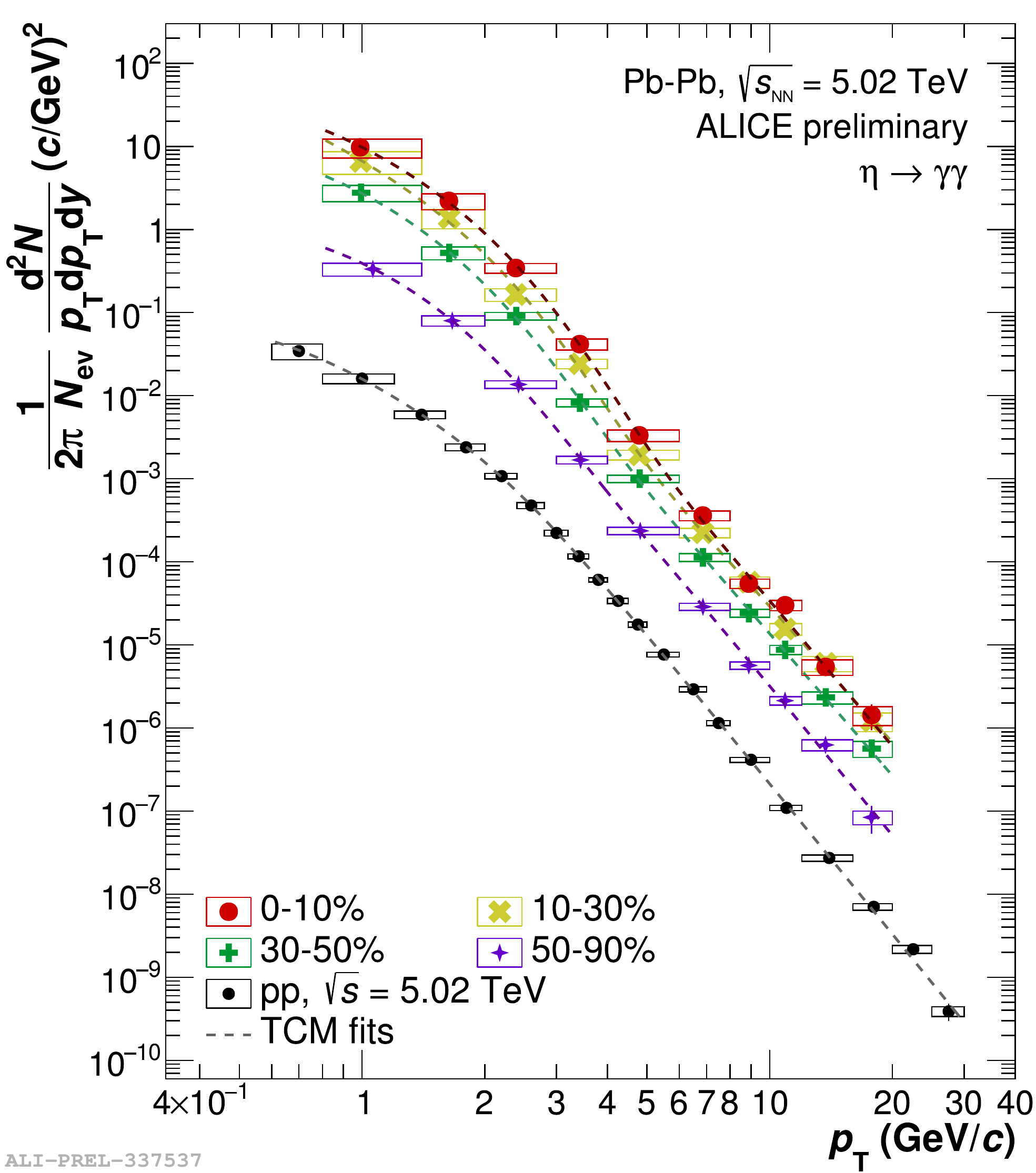}
	\caption{The production of $\pi^{0}$'s (left) and $\eta$ (right) in pp and Pb--Pb collisions at \s$ = 5$ TeV, as measured with the ALICE detector. The mesons are measured via the two photon decay channel, and the photons are reconstructed using the photon conversion method and the calorimeters PHOS and EMCal.}
	\label{pp_PbPb_production}
\end{figure}

\begin{figure}
	\centering
	\includegraphics[width=0.49\textwidth]{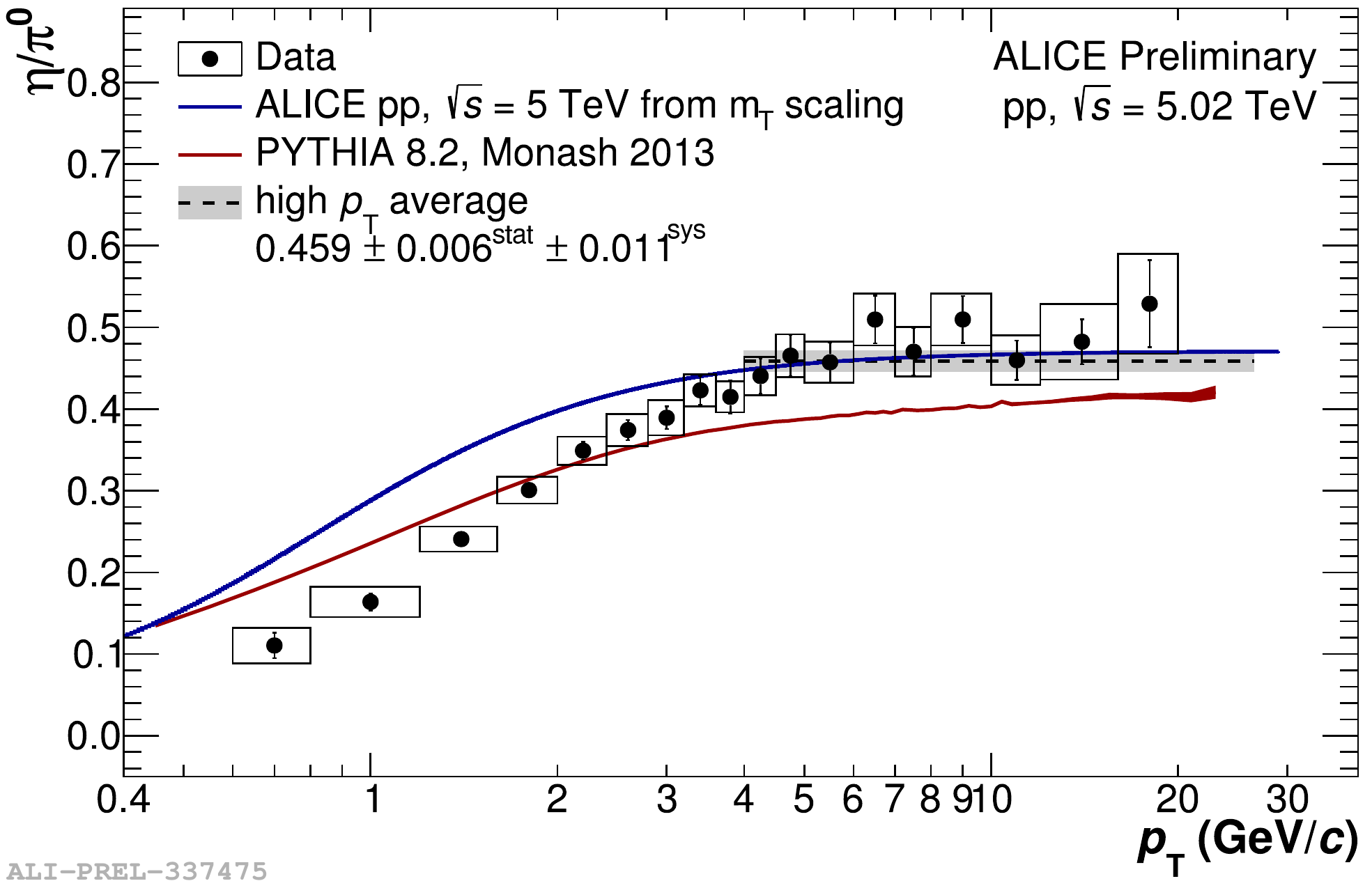}
	\includegraphics[width=0.49\textwidth]{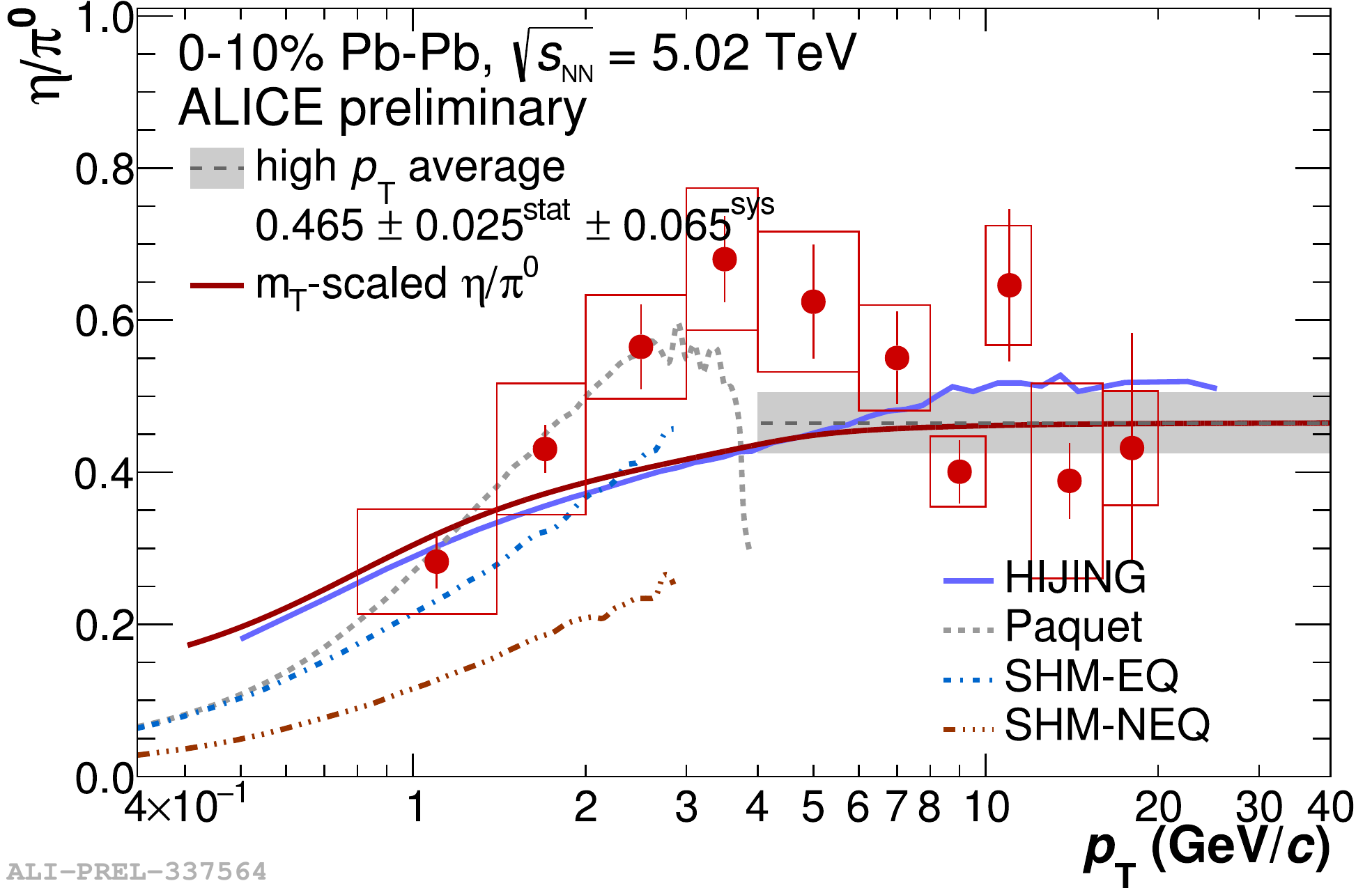}
	\caption{The $\eta / \pi^{0}$ ratio for pp collisions at \s$ = 5$ TeV (left), and \Pb collisions at \sNN$ = 5.02$ TeV (right).}
	\label{etapi0_ratio}
\end{figure}

\section{Results}
\subsection{Neutral meson production in \pp and \Pb collisions}

Figure \ref{pp_PbPb_production} shows the invariant yield of neutral pions (left) and $\eta$ mesons (right) as function of transverse momentum in pp collisions at \s$ = 5$ TeV and \Pb collisions at \sNN$ = 5$ TeV. It is measured using photons reconstructed with PCM and the calorimeters PHOS and EMCal.
Furthermore, Fig. \ref{etapi0_ratio} shows the $\eta / \pi^{0}$ ratio for pp collisions at \s$ = 5$ TeV (left), and \Pb collisions at \sNN$ = 5.02$ TeV (right). These results are compared to several model calculations. In \pp collisions, $m_{T}$ scaling ~\cite{Bourquin:1976} does not describe the data at low $\pT$, while PYTHIA~\cite{Pythia:2014} only approximately reproduces the ratio. In \Pb collisions, a hint of an enhancement of $\eta / \pi^{0}$ ratio around 3 GeV/$c$ with respect to HIJING and theoretical calculations can be seen, which could be attributed to radial flow effects.

\subsection{Nuclear modification of neutral mesons}

\begin{figure}
	\centering
	\includegraphics[width=0.49\textwidth]{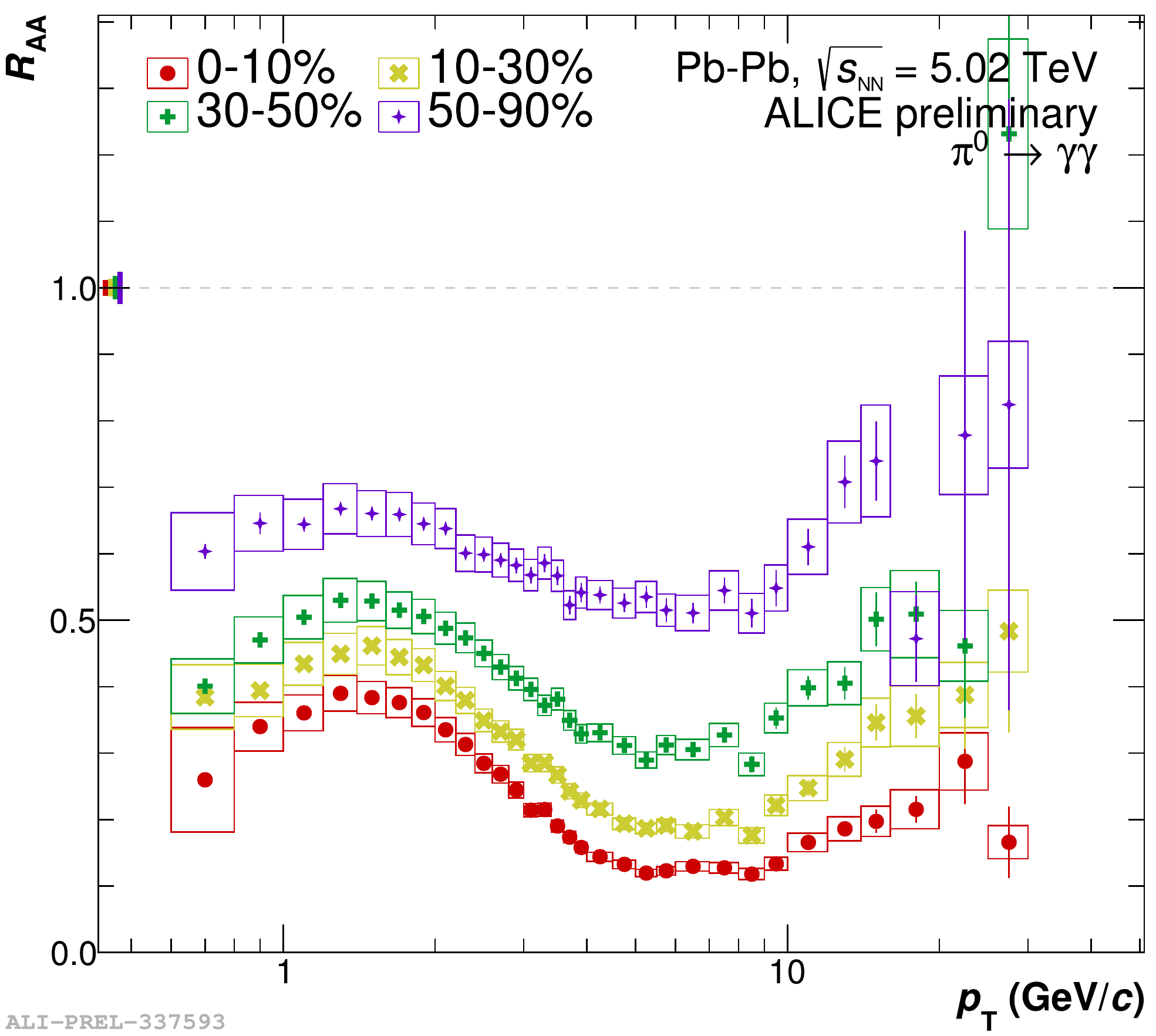}
	\includegraphics[width=0.49\textwidth]{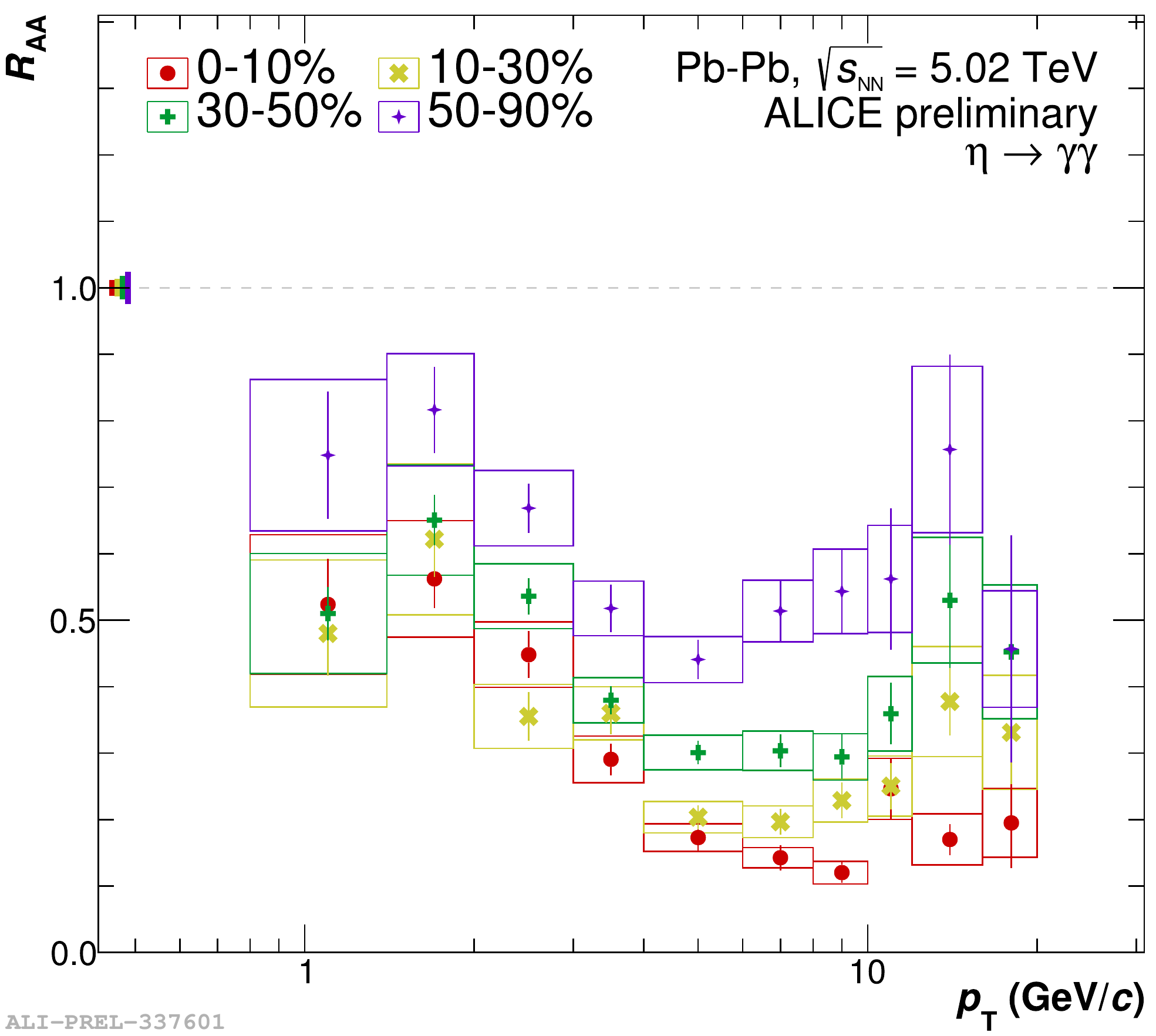}
	\caption{The nuclear modification factor R$_{\mathrm{AA}}$ of neutral pions (left) and $\eta$ mesons (right) in Pb--Pb collisions at \sNN$ = 5$ TeV.}
	\label{Fig:PbPb_pi0}
\end{figure}

Figure \ref{Fig:PbPb_pi0} shows the nuclear modification factor $R_{\mathrm{AA}} = (dN^{\mathrm{AA}}/d p_{\mathrm{T}})/(<T_{\mathrm{AA}}>d\sigma^{\mathrm{pp}}/d p_{\mathrm{T}})$ of neutral pions (left) and $\eta$ mesons (right) in Pb--Pb collisions at \sNN$ = 5$ TeV. A clear suppression is observed for both neutral mesons, where central \Pb collisions show more suppression than peripheral ones. The $R_{\mathrm{AA}}$ is similar for both mesons, despite the different quark content. Furthermore, it is the first time that the $\eta$ meson is measured in such a large range of $\pT$ and to such high precision, which is crucial in understanding the background present in direct photon and di-lepton measurements.

\section{Conclusion}

The neutral meson invariant yield in pp and \Pb collisions has been measured with the ALICE detector, utilising all the available photon reconstruction methods and combining the neutral meson measurements, improving the precision and extending the $\pT$ range. 

The $\eta / \pi^{0}$ ratio was measured in pp and AA collisions. In pp it shows the universal behavior, independent of collision energy. At high $\pT$ the ratio is by construction reproduced with $m_{T}$ scaling, and PYTHIA only approximately reproduces the $\pT$ dependence. In AA collisions the modifications to the $\eta / \pi^{0}$ ratio are characteristic for the presence of radial flow. In addition, the nuclear modification factors $R_{\mathrm{AA}}$ of the $\pi^{0}$ and $\eta$ mesons are calculated for \Pb collisions, using the respective invariant yield in pp collisions, and are found to be similar despite the different quark content.

\bibliographystyle{JHEP}
\bibliography{biblio.bib}











\end{document}